\documentclass[10pt,conference]{IEEEtran}
\usepackage[a4paper,left=1.2cm,right=1.2cm,top=1.8cm,bottom=3.7cm]{geometry}
\usepackage{cite}
\usepackage{amsmath,amssymb,amsfonts}
\usepackage{algorithmic}
\usepackage[ruled,vlined,linesnumbered]{algorithm2e}
\SetAlgorithmName{Algorithm}{Algorithm}{List of Algorithms}
\usepackage{graphicx}
\usepackage{stmaryrd}
\usepackage{xcolor}
\usepackage{array}
\usepackage{commath}
\usepackage{sidecap}
\usepackage{stfloats}
\usepackage{float}
\usepackage{tabularx, boldline}
\usepackage{rotating,booktabs,multirow}
\usepackage{mathtools}
\usepackage{flexisym}
\usepackage{mathtools}
\usepackage{amsmath}
\usepackage{graphicx}
\usepackage{tabularx, boldline}
\usepackage{rotating,booktabs,multirow}
\usepackage{enumerate}
\usepackage{etoolbox}
\usepackage[]{nomencl}   
    \makenomenclature
\usepackage{xcolor}
\usepackage{array}
\usepackage{commath}
\usepackage{sidecap}
\usepackage{stfloats}
\usepackage{tabularx, boldline}
\usepackage{rotating,booktabs,multirow}
\usepackage{mathtools}
\usepackage{flexisym}
\usepackage{breqn}
\usepackage{url}
\usepackage{adjustbox}
\usepackage{makecell}

\usepackage{textcomp}
\usepackage{amsmath}
\usepackage{txfonts}
\usepackage{graphicx}
\usepackage{acronym}
\usepackage{tabularx, boldline}
\usepackage{rotating,booktabs,multirow}
\usepackage{enumerate}
\usepackage{etoolbox}
\usepackage[]{nomencl}   
    \makenomenclature

\def\E{\mathbb{E}} 
\ifCLASSINFOpdf
\else
\fi

\begin{document}

\title{Learning Sparse Privacy-Preserving Representations for Smart Meters Data}
\author{\IEEEauthorblockN{Mohammadhadi Shateri}
\IEEEauthorblockA{\textit{McGill University} \\  Montreal, Canada}
\and
\IEEEauthorblockN{Francisco Messina}
\IEEEauthorblockA{\textit{Universidad de Buenos Aires} \\  Buenos Aires, Argentina}
\and
\IEEEauthorblockN{Pablo Piantanida}
\IEEEauthorblockA{Laboratoire des Signaux et Systèmes (L2S)\\ CentraleSup\'elec CNRS Universit\'e Paris-Saclay\\
 Gif-sur-Yvette, France}
\and
\IEEEauthorblockN{Fabrice Labeau}
\IEEEauthorblockA{\textit{McGill University} \\  Montreal, Canada}}

\maketitle

\begin{abstract}
Fine-grained Smart Meters (SMs) data recording and communication has enabled several features of Smart Grids (SGs) such as power quality monitoring, load forecasting, fault detection, and so on. In addition, it has benefited the users by giving them more control over their electricity consumption. However, it is well-known that it also discloses sensitive information about the users, i.e., an attacker can infer users' private information by analyzing the SMs data. In this study, we propose a privacy-preserving approach based on non-uniform down-sampling of SMs data. We formulate this as the problem of learning a sparse representation of SMs data with minimum information leakage and maximum utility. The architecture is composed of a releaser, which is a recurrent neural network (RNN), that is trained to generate the sparse representation by masking the SMs data, and an utility and adversary networks (also RNNs), which help the releaser to minimize the leakage of information about the private attribute, while keeping the reconstruction error of the SMs data minimum (i.e., maximum utility). The performance of the proposed technique is assessed based on actual SMs data and compared with uniform down-sampling, random (non-uniform) down-sampling, as well as the state-of-the-art in privacy-preserving methods using a data manipulation approach. It is shown that our method performs better in terms of the privacy-utility trade-off while releasing much less data, thus also being more efficient.
\end{abstract}

\IEEEpeerreviewmaketitle

\section{Introduction}
The Smart Grid (SG) aims at improving the efficiency, reliability, and security of the electric power systems by using intelligent transmission, control, and distribution networks~\cite{kabalci2019introduction}. One of the main components of the SG are smart meters (SMs), which are devices that enable data exchange between users and Utility Provider (UP) by recording the electricity data of the consumers~\cite{5403159}. This fine-grained electricity consumption data is used for billing, load forecasting, energy theft detection, and several other applications for improving the grid operation. However, the SMs data can be eavesdropped or be shared with a third-party, which can potentially infer sensitive information about users, including the behavioural patterns or even the type of appliances used in the dwelling~\cite{giaconi2018privacy,6003811}.

There are several privacy-preserving techniques that have been proposed to address the SMs data privacy issue, which can be divided in two main categories: Data Manipulation (DM) and Demand Load Shaping (DLS). The DM approaches modify the SMs using techniques such as data obfuscation/perturbation, anonymization, down-sampling, etc. ~\cite{efthymiou2010smart,sankar2013smart, yang2016evaluation, shateri2020privacy, barbarosa2016,tripathy2019privacy, shateri2019deep, shateri2020a,9302948}. Many of the most recent studies of this family, incorporated information theoretic measures such as Mutual Information (MI) or Directed Information (DI) to model the amount of information leaked about the sensitive attributes and used Machine Learning (ML) algorithms for their implementation. On the other hand, the DLS approaches use physical resources such as rechargeable batteries, electric vehicles, and even renewable energy resources, to shape the users' power consumption to mask the sensitive patterns ~\cite{kalogridis2010privacy,yao2013privacy,tan2013increasing,gomez2014smart,zhang2016cost,giaconi2017smart,li2018information,erdemir2019privacy,giaconi2017optimal,sun2017smart, shateri9248831, shateri2020privacycost}. Recently, Reinforcement Learning (RL) and Deep Reinforcement Learning (DRL) methods have been used to tackle this problem, showing good performance against strong ML based attackers. 

One of the most simple and naive methods in the DM family is that of down-sampling. Although this is a straightforward and efficient mechanism that can reduce the stress on the communication channel and storage requirements, by communicating SMs data with lower rate, it has received less attention than other techniques. The motivation for this method is based on the fact that high granularity or temporal resolution of data can tremendously improve the accuracy of the electricity consumption disaggregation methods~\cite{huchtkoetter2020impact}. Therefore, by down-sampling the data, the performance of the disaggregation algorithms can be controlled, which means in turn that less sensitive information is shared with third-parties or UPs ~\cite{cardenas2012privacy,mashima2015authenticated,eibl2015,zhang2020privacy}. In the literature, however, there have been more efforts on motivating and analyzing down-sampling than presenting a comprehensive way for getting the most use out of this technique. In This work, we adopt a non-uniform down-sampling approach using deep neural networks for its implementation. Concretely, we pose the problem as that of learning a sparse representation of SMs data by decimating some of its samples with a learned probability distribution. Our framework includes three deep recurrent neural networks (RNNs): a releaser, a utility, and an adversary. The releaser is trained to generate a sparse representation of the SMs data by getting feedback from the utility and adversary networks regarding the reconstruction error of the original SM data, and privacy of the representation, respectively. Following our earlier study~\cite{shateri2020a}, DI between the sensitive data and its estimation (by the adversary network) is used as privacy measure and the mean squared error between SM data and its reconstructed version (by the utility network) is considered as the utility measure. Finally, the performance of the presented framework is tested using actual SM data and compared empirically with a state-of-the-art method ~\cite{shateri2020a}, uniform down-sampling, and random (non-uniform) down-sampling in terms of privacy-utility trade-off, average released data rate, and the leakage of information about the sensitive attribute. To the best of our knowledge, this is the first work where a down-sampling privacy-preserving approach is implemented using deep neural networks for SM data.

The rest of the paper is organized as follows. In Section \ref{sec:Prob_formul}, the problem formulation of the SMs privacy-utility based on down-sampling is developed in relation with sparse representation of the SMs. Details of implementation of the proposed technique using deep RNNs are given in Section \ref{PPM_Implementation}. Empirical results for our method and comparisons with other approaches are presented and discussed in Section \ref{sec:results}. Finally, some concluding remarks are presented in Section \ref{sec:conclusion}.

\section{Problem Formulation}\label{sec:Prob_formul}

\begin{figure}[htbp!]
	\centering
	\includegraphics[width=0.95\linewidth]{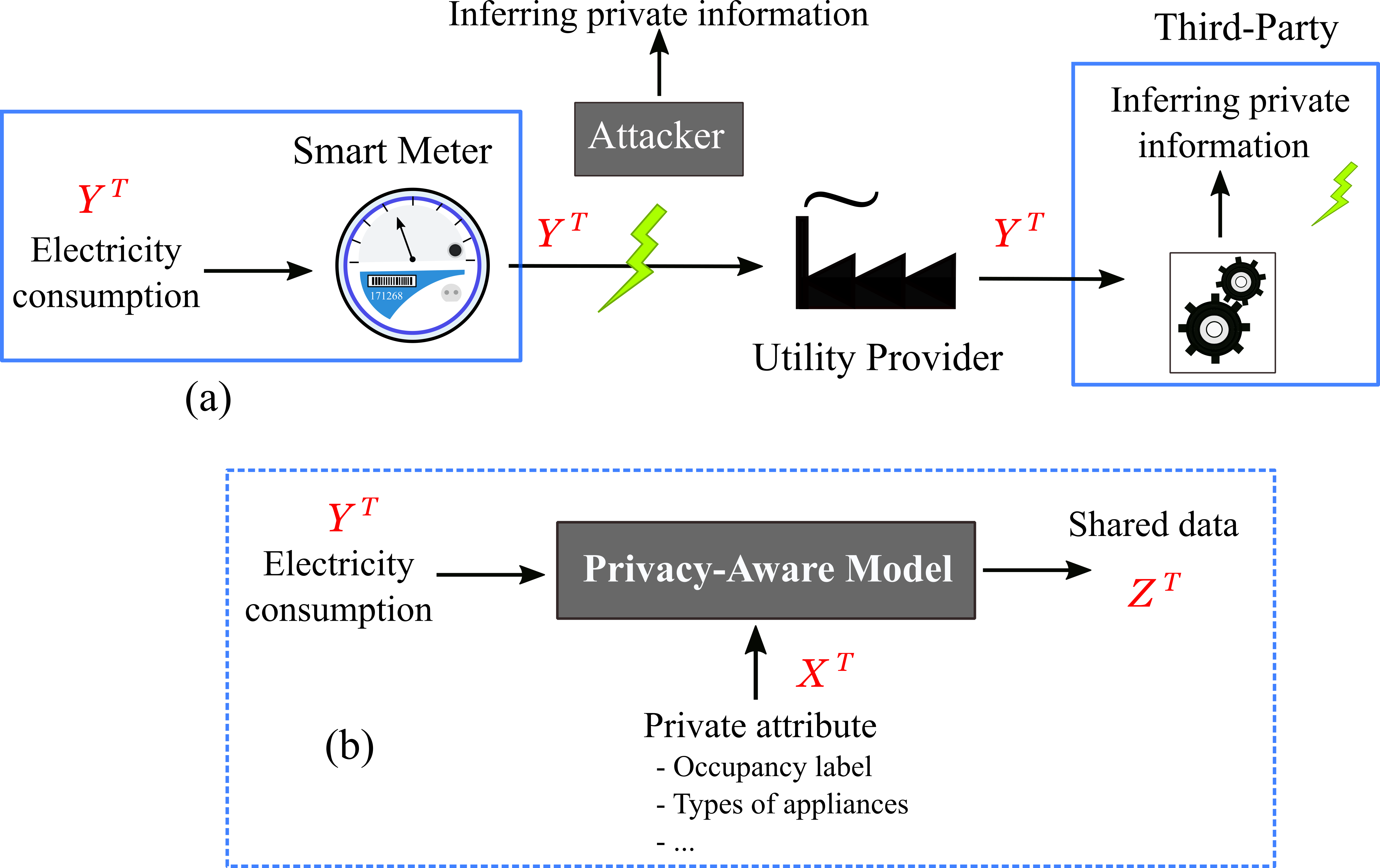}
	\caption{(a) Smart meter data are communicated with the utility provider in real-time and could be attacked by either an eavesdropper or a third-party. (b) Sanitizing the smart meter data using a privacy-aware model before communicating it with the utility provider.}
	\label{fig:privacy-aware-model}
\end{figure}

Consider the electricity consumption of a household (useful data), denoted as a time series sequence $Y^T = \{Y_t\}_{t=1}^{T}$, that is recorded by SMs and needs to be communicated in real-time to the UP. To avoid violating the users' privacy, a privacy-preserving mechanism generates a representation of $Y^T$, denoted as $Z^{T} = \{Z_t\}_{t=1}^{T}$, which attempts to preserve the utility of $Y^T$ while leaking minimum information about a private attribute $X^T = \{X_t\}_{t=1}^{T}$ that needs to be hidden. Therefore, instead of the actual SM data $Y^{T}$, its new representation $Z^{T}$ would be released and shared with the UP. The scheme is shown in Fig.\ref{fig:privacy-aware-model}.

In this work, the mechanism used by the privacy-preserving system is based on reducing the granularity of the SM data, i.e. down-sampling the data. However, unlike common down-sampling, we propose a more sophisticated non-uniform down-sampling by deciding whether to release the data at each time instant $t$ or not. Specifically, the release mechanism $\mathcal{M}$ is as follows:
\begin{equation} \label{Mechanism_sparse}
    \mathcal{M}:\; W^{t}\rightarrow Z_{t} = \left\{
\begin{array}{lr}
	Y_t \hspace{0.5cm} \text{w.p. }\; q_t(W^{t})\\
	\\
	0\; \hspace{0.5cm} \text{w.p. }\; 1- q_t(W^{t})\\
\end{array}
\right.
\end{equation}
where $t\in\{1,\dots,T\}$, $q_t(.)$ determines the chance of releasing sample $Y_t$ at time instant $t$, and $W_t$ is the observed data or input to by privacy-preserving system at time $t$ (which uses both $X_t$ and $Y_t$, see Algorithm \ref{AL_TD} for details). Notice that $\mathcal{M}$ produces a sparse representation of the data, which is stochastic and a function of the input $W^T$.

Following the study~\cite{shateri2020a}, we use the DI $I(X^T\rightarrow \hat{X}^T)$ between the private attribute $X^T$ and its estimation by a worst-case adversary $\hat{X}^T$ as the privacy measure. On the other hand, the utility is measured based on the expected distortion between $Y^T$ and its best reconstruction $\hat{Y}^T$ based on $Z^T$. Therefore, the problem of finding the optimal sparse representation following the mechanism $\mathcal{M}$ in \eqref{Mechanism_sparse}, which leaks minimum information about sensitive attribute while keep the utility of the data, can be formulated as follows:
\begin{align} \label{eq:DI_optimization} \underset{q^{T}}{\text{min}}  \;  \frac{1}{T} I\left(X^T\rightarrow \hat{X}^T\right) \quad \text{s.t.} \quad \frac{\mathbb{E}\left[||Y^T-\hat{Y}^T||_2^2\right]}{T} \le \varepsilon, \end{align}
where $\|\cdot \|_2$ is the euclidean norm, and $\varepsilon>0$ is the maximum tolerance on the expected reconstruction error.

\section{Privacy-preserving Framework and Implementation}\label{PPM_Implementation}

\begin{figure}[htbp!]
    \centering
    \includegraphics[width=0.95\linewidth]{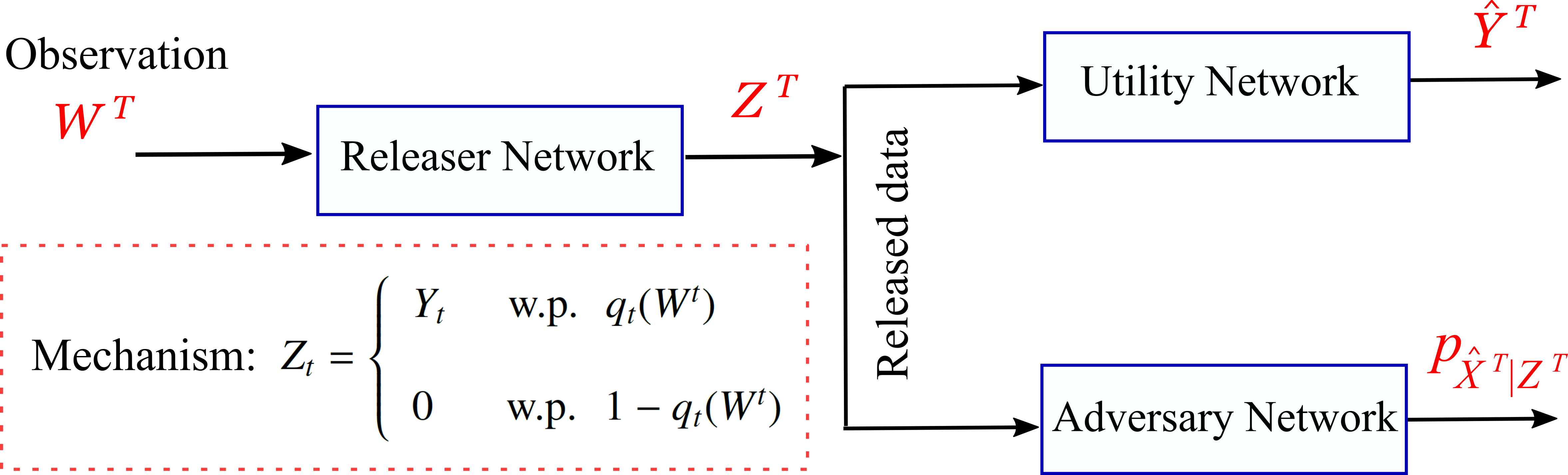}
    \caption{Privacy-preserving model framework for sharing/releasing the SMs data.} 
    \label{fig:PA_structure}
\end{figure}

The general framework for designing the privacy-preserving system is shown in Fig.~\ref{fig:PA_structure}. Notice that, in addition to a releaser network, two more networks named as utility and adversary are included. On the one hand, the releaser keeps the reconstruction error minimum by getting feedback from the utility network, which estimates the useful data. On the other hand, the adversary network, which estimates the sensitive attribute from the released data, provides feedback for the releaser network to measure the leakage of information about the sensitive data. This process continues until all networks converge.

\begin{figure}
    \centering
    \includegraphics[width=0.95\linewidth]{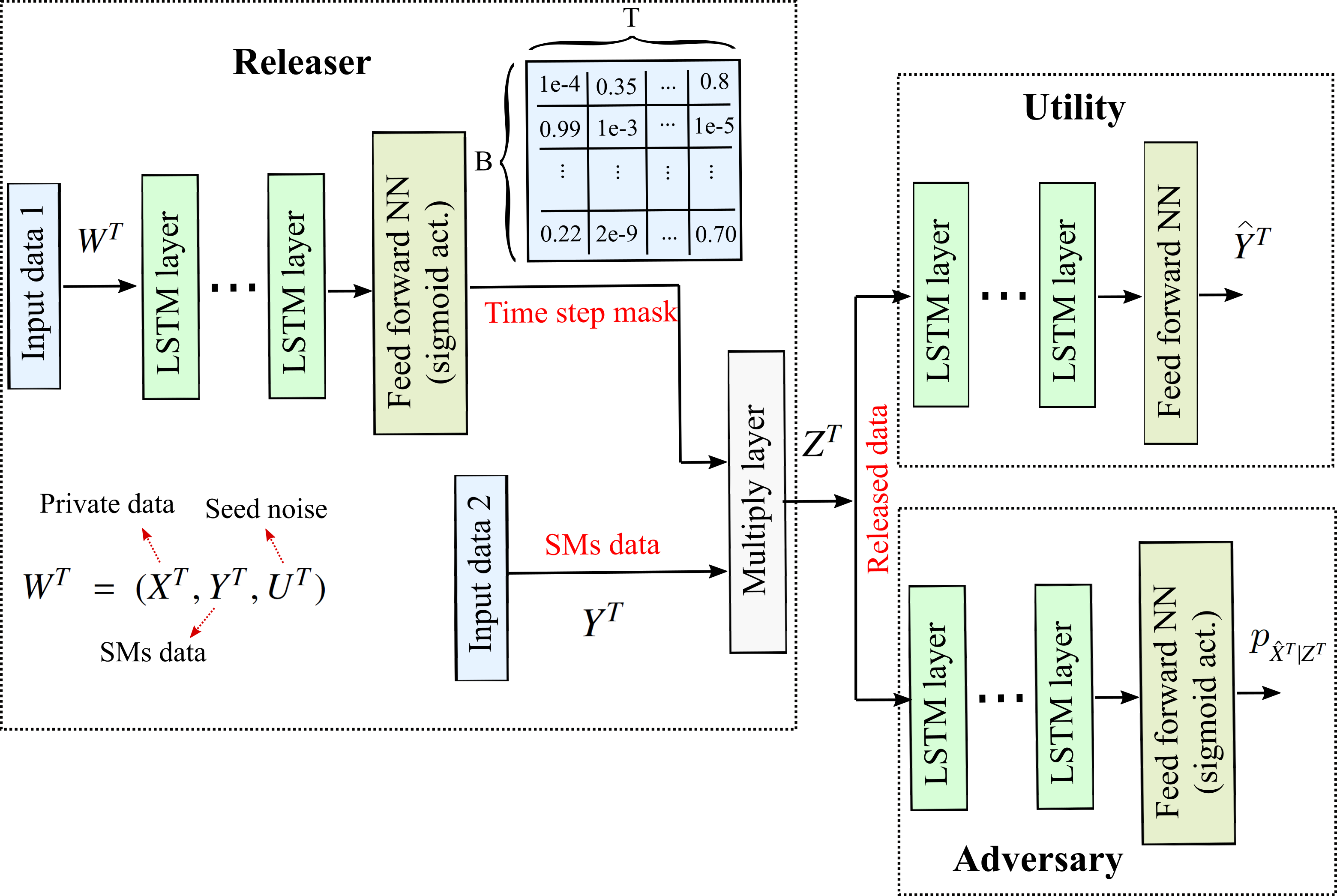}
    \caption{Privacy-preserving framework implemented based on Long-Short Term Memory (LSTM) layers. The seed noise is generated according to a Uniform distribution, i.e. $U_t\sim U[0,1]$.}
    \label{fig:PA_LSTM}
\end{figure}

In the following, we describe how to implement the privacy-preserving framework of Fig.\ref{fig:PA_structure} using deep neural networks. To this end, the sparse representation $Z^T$ is considered as a masked version of the $Y^T$ as $Z^T = Y^T\circ M^T$ where $\circ$ is the Hadamard product (or element-wise product) and $M^T$ is a $0-1$ mask. In the training phase, the releaser network would learn a soft mask with elements $q_t$ where $0\leq q_t \leq 1$. This can be done using the sigmoid function $\sigma(z) \coloneqq 1 / (1 + \exp(-z))$. Fig.\ref{fig:PA_LSTM} shows the privacy-preserving framework implemented using RNNs to model the temporal correlation in the data.
We now define the loss functions for training each of the networks. On the one hand, since the goal of the utility network is to recover the actual SMs data, we use the expected distortion as its loss function:
\begin{equation} \label{eq:DI_utility_loss} \mathcal{L}_{\mathcal{U}}(\psi) \coloneqq  \frac{1}{T} \E\left[||Y^T-\hat{Y}^T||_2^2\right], \end{equation}
where $\psi$ are the parameters of the utility network. Let us assume that $X_t$ is a discrete random variable, so the adversary act as a classifier and we use the cross-entropy as its loss function:
\begin{equation} \label{eq:DI_adversary_loss} \mathcal{L}_{\mathcal{A}}(\phi) \coloneqq \frac{1}{T} \sum_{t=1}^{T} \E\left[-\log p_{\hat{X}_t|Z^t}(X_t|Z^t) \right], \end{equation}
where $\phi$ are the parameters of the utility and adversary networks. Finally, the loss functions of the releaser network can be determined based on the optimization problem \eqref{eq:DI_optimization}. Since the DI is intractable, we use the following upper bound~\cite{shateri2020a}:
\begin{align} \label{eq:di_upper_bound} I\left(X^T\rightarrow \hat{X}^T\right) \leq T \log|\mathcal{X}| - \sum_{t=1}^{T}H(\hat{X}_t|Z^t). \end{align}
Therefore, by substituting this bound as a surrogate of the DI, the following loss function is obtained:
\begin{equation} \label{eq:DI_releaser_loss} \mathcal{L}_{\mathcal{R}}(\theta, \phi,\psi,\lambda) \coloneqq  \frac{1}{T} \E\left[||Y^T-\hat{Y}^T||_2^2\right] - \frac{\lambda}{T}\sum_{t=1}^{T} H\big(\hat{X}_t|Z^t\big), \end{equation}
where $\lambda$ controls the privacy-utility trade-off. The complete training algorithm for this privacy-aware framework is presented in detail in Algorithm \ref{AL_TD}.

\begin{algorithm}
    \footnotesize
    \algsetup{linenosize=\tiny}
	\caption{Privacy-preserving model based on down-sampling. Batch size $B$, seed noise dimension $m$, number of steps to apply to the Adversary $k$, and $\ell_2$ regularization parameter $\beta$ are hyperparameters.}
	\label{AL_TD}
	\begin{algorithmic}[1]
	\FOR {number of training iterations}
	    \FOR {$k$ steps}
		\STATE Sample minibatch of $B$ examples $\{ w^{(b)T} = (x^{(b)T},y^{(b)T},u^{(b)T})\}_{b=1}^B$.
		\STATE Generate released data $\{ z^{(b)T}\}_{b=1}^B$ as the Hadamard product of soft mask $\{ M^{(b)T}\}_{b=1}^B$ and data $\{ y^{(b)T}\}_{b=1}^B$.
		\STATE Compute the gradient of $\mathcal{L}_{\mathcal{A}}(\phi)$, approximated empirically for minibatch, with respect to $\phi$ and update $\phi$ by applying the RMSprop optimizer ~\cite{hinton2012neural}.
		\ENDFOR
		
		\STATE Sample minibatch of $B$ examples $\{ w^{(b)T}\}_{b=1}^B$ and generate $\{ z^{(b)T}\}_{b=1}^B$.
		\STATE Compute the gradient of $\mathcal{L}_{\mathcal{U}}(\psi)$, approximated empirically for minibatch, with respect to $\psi$ and update $\psi$ by applying the RMSprop optimizer.
		\STATE Compute the gradient of $\mathcal{L}_{\mathcal{R}}(\theta,\phi,\psi,\lambda)$, approximated empirically for minibatch, with respect to $\theta$.
		\STATE Use $\textrm{Ridge}(L_2)$ regularization \cite{hastie2005elements} with value $\beta$ and update $\theta$ by applying RMSprop optimizer.
		\ENDFOR
	\end{algorithmic}
\end{algorithm}

\begin{figure}
    \centering
    \includegraphics[width=0.80\linewidth]{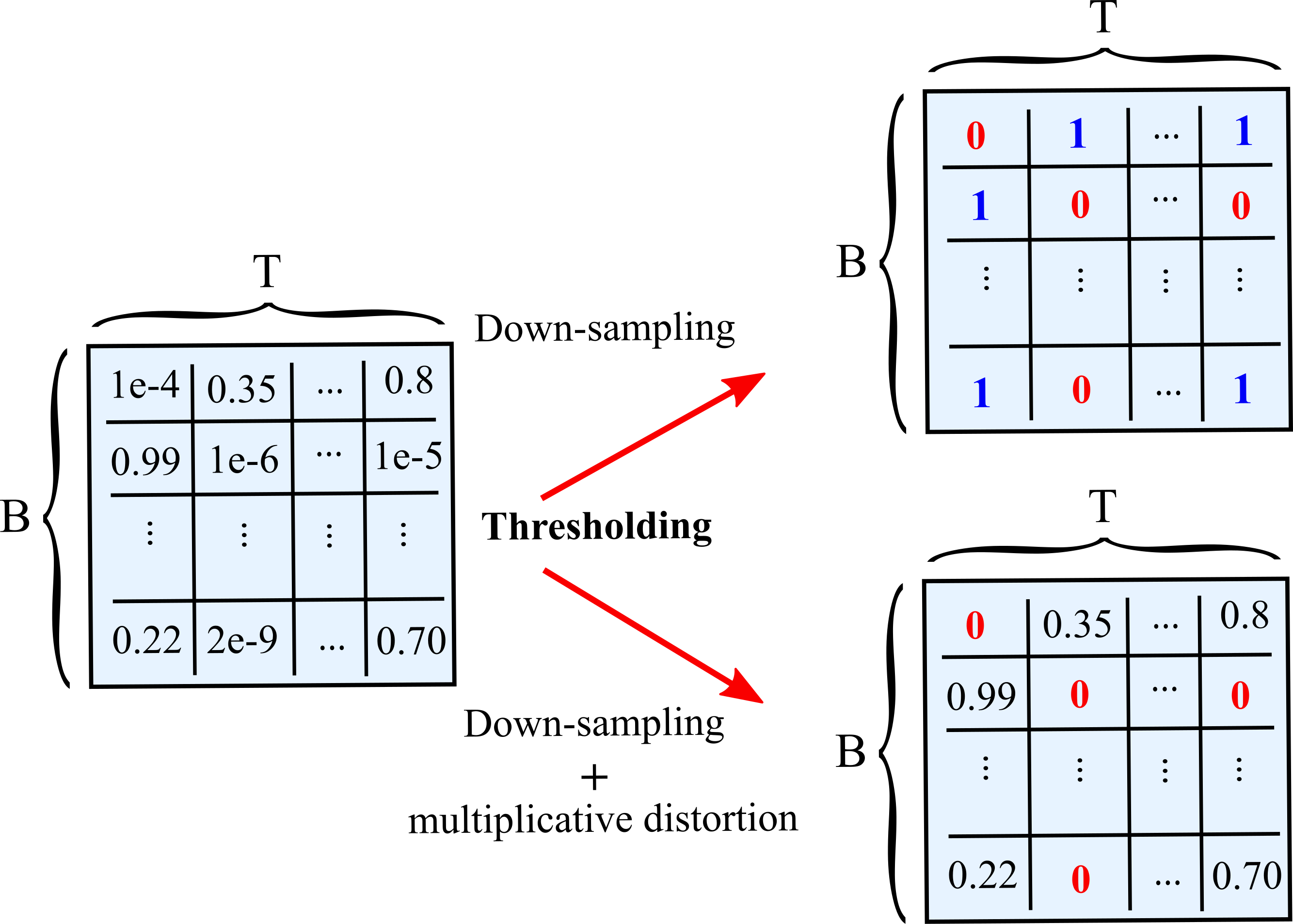}
    \caption{Soft mask thresholding in the testing phase of the privacy-preserving model. The level of threshold is a hyperparameter.}
    \label{fig:Deci_test}
\end{figure}

After training the privacy-preserving framework in Fig.\ref{fig:PA_LSTM} using Algorithm \ref{AL_TD}, in the testing phase, the soft mask can be treated in two ways by changing the thresholding operation: considering a $0-1$ mask or a zero and non-zero mask (see Fig.\ref{fig:Deci_test}). In the former case, privacy is provided just using the down-sampling mechanism, i.e., removing some time instances for hiding sensitive information, while in the later case both down-sampling and multiplicative distortion mechanisms are used.  

\section{Results and Discussion} \label{sec:results}

In this study, the performance of the proposed model is assessed based on the Electricity Consumption $\&$ Occupancy (ECO) dataset. The ECO dataset collected and proposed by~\cite{beckel2014eco} includes 1 Hz power consumption of five houses in Switzerland labeled with the occupancy status of the dwellings. These occupancy labels are considered as a private attribute. Therefore, the goal of the releaser network is to generate a sparse representation from which an adversary cannot infer the occupancy labels while the utility network can use the released data to recover the actual power consumption. In this setting, the releaser and utility are regression networks while the adversary is a binary classifier. To make the proposed model comparable with the state-of-the-art method such as the one proposed in~\cite{shateri2020a}, the dataset is re-sampled every one hour and daily samples (with length $T=24$) are considered. The total 11225 sample sequences are split into train and test dataset with the ratio $85:15$, and $10\%$ of the training dataset is used as the validation dataset to tune the hyperparameters of the model. The proposed smart down-sampling method (with and without multiplicative distortion) is compared with uniform down-sampling, random (non-uniform) down-sampling, and the additive distortion approach proposed in~\cite{shateri2020a}. The architectures of the networks used for each method are presented in Table~\ref{tab:models}. Notice that, unlike the smart down-sampling, for the uniform down-sampling and random down-sampling, since there are no parameters to be learned, the sparse representation is generated independently of the utility network. It should be noted that the releaser for the random method would be a non-uniform down-sampling of the SMs data where the decimated time instances are selected randomly.

\begin{table*}[htbp]
	\centering
	\caption{Networks architectures and hyperparameters values for the privacy-aware models. }
	\begin{adjustbox}{width=0.85\textwidth}
		\begin{tabular}{c c c c c c c c c}
			\toprule
			\textbf{\makecell{}}&
			\textbf{\makecell{Releaser}}& \textbf{\makecell{Adversary}}&
			\textbf{\makecell{Utility}}&
			\textbf{\makecell{Attacker}}& \textbf{\makecell{$B$}}& \textbf{\makecell{$k$}}& \textbf{\makecell{$m$}}\\ 
			\textbf{\makecell{}}&\makecell{$4$ LSTM layers each\\with 64 cells and $\beta=1.5$} &\makecell{$2$ LSTM layers each\\ with 32 cells}&\makecell{$3$ LSTM layers each\\ with 48 cells}& \makecell{$3$ LSTM layers each\\ with 32 cells}&128 &4&8 \\
		    \midrule[0.5pt]
		    \textbf{\makecell{Smart Down-Sampling}}&\makecell{\checkmark}&\makecell{\checkmark}&\makecell{\checkmark}&\makecell{\checkmark}&\makecell{\checkmark}&\makecell{\checkmark}&\makecell{\checkmark}\\
		    \midrule[0.5pt]
		    \textbf{\makecell{Smart Down-Sampling\\ and Multiplicative Distortion}}&\makecell{\checkmark}&\makecell{\checkmark}&\makecell{\checkmark}&\makecell{\checkmark}&\makecell{\checkmark}&\makecell{\checkmark}&\makecell{\checkmark}\\
		    \midrule[0.5pt]
		    \textbf{\makecell{Uniform Down-Sampling*}}&\makecell{\textemdash}&\makecell{\textemdash}&\makecell{\checkmark}&\makecell{\checkmark}&\makecell{\textemdash}&\makecell{\textemdash}&\makecell{\textemdash}\\
		    \midrule[0.5pt]
		    \textbf{\makecell{Random Down-Sampling}}&\makecell{\textemdash}&\makecell{\textemdash}&\makecell{\checkmark}&\makecell{\checkmark}&\makecell{\textemdash}&\makecell{\textemdash}&\makecell{\textemdash}\\
		    \midrule[0.5pt]
		    \textbf{\makecell{Additive Distortion~\cite{shateri2020a}}}&\makecell{\checkmark}&\makecell{\checkmark}&\makecell{\textemdash}&\makecell{\checkmark}&\makecell{\checkmark}&\makecell{\checkmark}&\makecell{\checkmark}\\
			\midrule
			\bottomrule
			\multicolumn{8}{l}{\small * To generate the released data, the \textit{resample} function of Matlab is used where a FIR Antialiasing Lowpass Filter is applied to the data.} \\
		\end{tabular}
	\end{adjustbox}
	\label{tab:models}
\end{table*}

\begin{figure}
    \centering
    \includegraphics[width=0.75\linewidth]{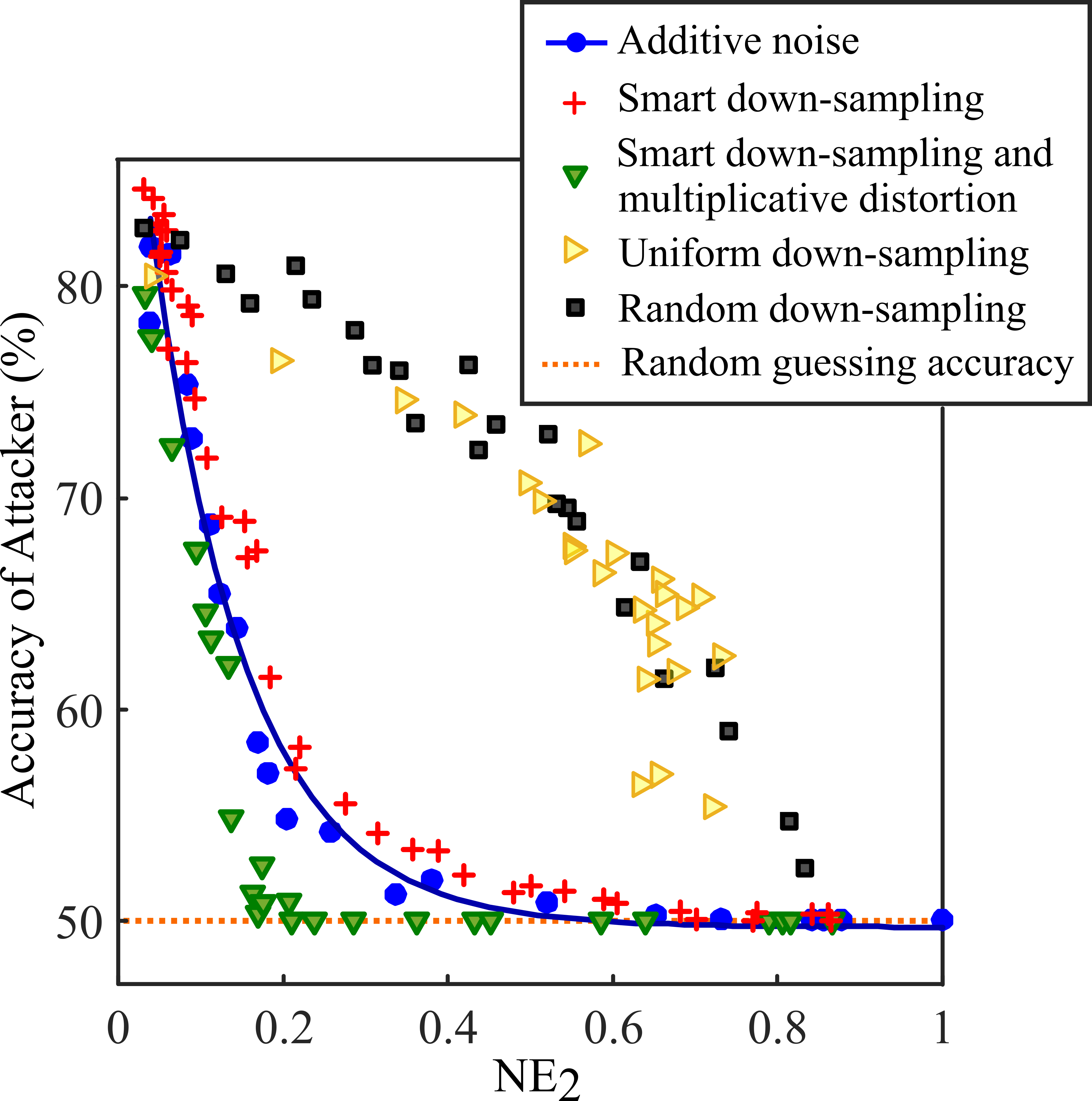}
    \caption{Privacy-utility trade-off of the proposed model compare with the uniform down-sampling, random (non-uniform) method, and additive noise approach~\cite{shateri2020a}. The result of the random method are averaged over five times random testing.}
    \label{fig:ALL_Models_Deci_test}
\end{figure}

As the first comparison, the privacy-utility of the models (on the test dataset) are assessed based on the performance of an attacker which is trained (by having access to the $(Z^T,X^T)$ training dataset, i.e., in a supervised manner) to infer the private attributes. It should be noted that since the attacker is a classifier, the performance of the attacker is evaluated using balanced accuracy, defined as follows~\cite{mosley2013balanced}:
\begin{equation}\label{eq:BA} \text{Balanced Accuracy} \coloneqq \frac{1}{2} \left( \frac{c_{11}}{c_{11}+c_{12}} + \frac{c_{22}}{c_{22}+c_{21}} \right), \end{equation}
where $c_{ij}$ is the fraction of samples of class $i$ classified as class $j$. The utility on the other hand, is measured based on the normalized square error, defined as follows:
\begin{equation}\label{eq:NE2} \text{NE}_2 \coloneqq \frac{\E\left[ \| Y^T - \hat{Y}^T \|_2 \right]}{\E\left[ \| Y^T \|_2 \right]}. \end{equation}
The privacy-utility trade-off for the different methods is shown in Fig.~\ref{fig:ALL_Models_Deci_test}. As expected, the smart down-sampling greatly outperforms both the uniform and the random down-sampling methods. In addition, the performance of the smart down-sampling is closely comparable with the additive noise approach, and smart down-sampling with multiplicative distortion approach actually outperforms the state-of-the-art.

\begin{figure}
    \centering
    \includegraphics[width=0.65\linewidth]{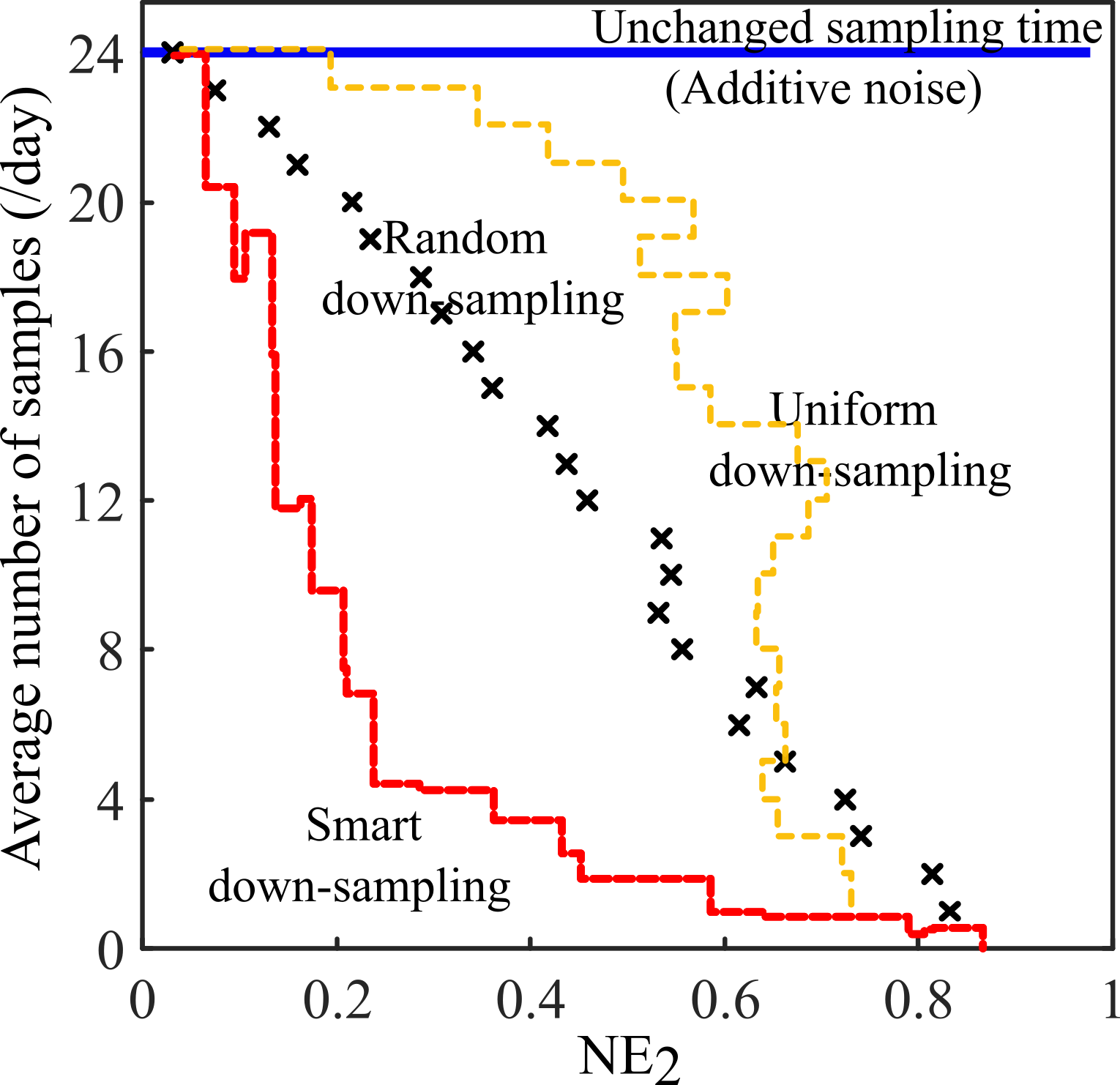}
    \caption{Averaged number of samples released daily versus the utility for the proposed models.}
    \label{fig:Samples_daily}
\end{figure}

As another comparison, we evaluate the average number of samples released daily (non-zero samples in down-sampling methods) by each scheme. Results are presented in Fig.~\ref{fig:Samples_daily}. Interestingly, looking at this figure and Fig.~\ref{fig:ALL_Models_Deci_test}, we can see that, for a fixed level of distortion, the smart down-sampling method outperforms the other methods in terms of privacy while also reducing the sample data rate.  

\begin{figure}
    \centering
    \includegraphics[width=0.8\linewidth]{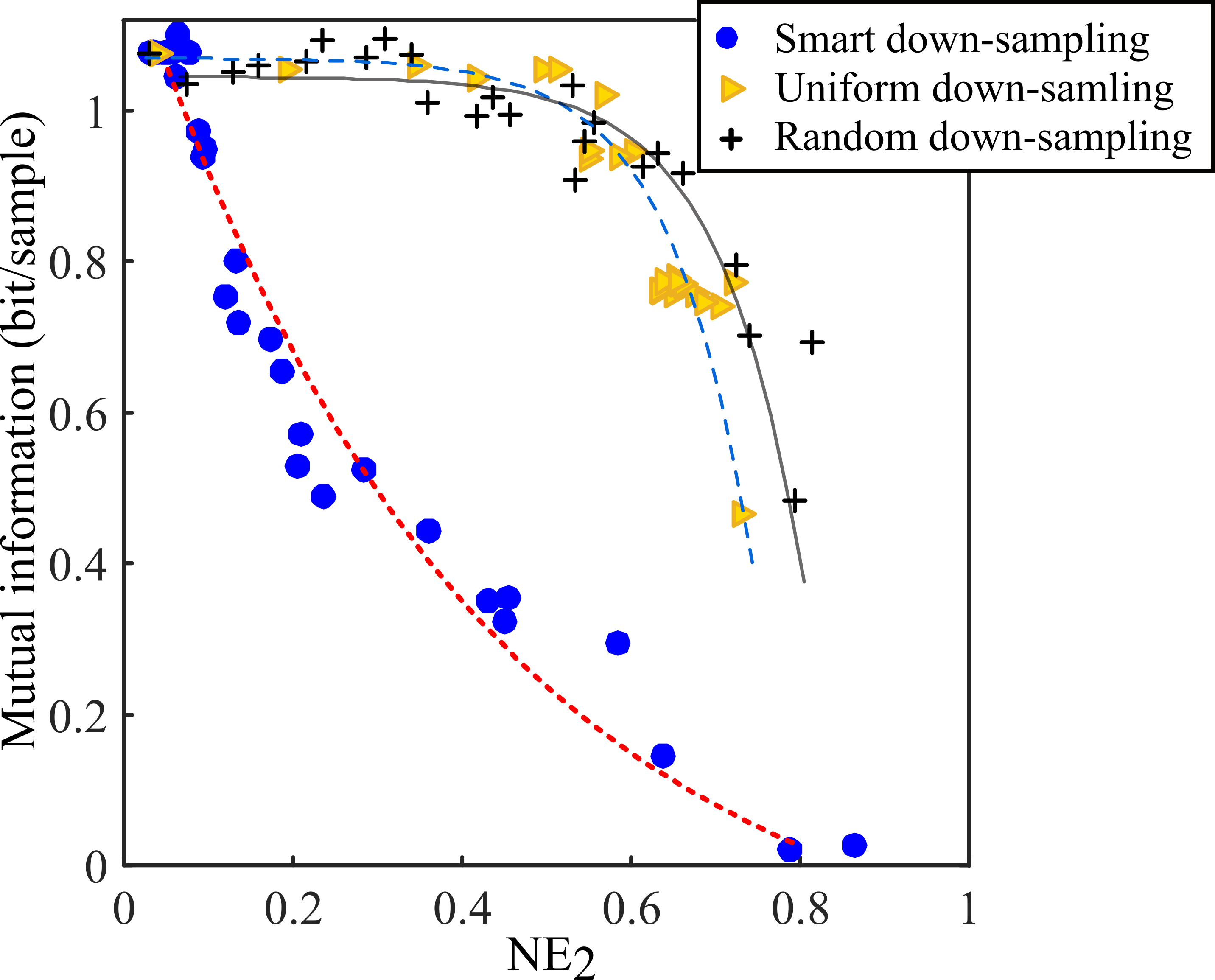}
    \caption{Information leakage about the users' private attribute (occupancy labels) from the shared data based on the KSG approximation of mutual information.}
    \label{fig:MI_Distortion}
\end{figure}

The general leakage of information about the users' private attribute (occupancy labels for this study) from the shared data is estimated by considering the mutual information between the occupancy labels $X^T$ and the shared data $Z^T$ as a function of the distortion (see Fig.~\ref{fig:MI_Distortion}). In this figure, the mutual information is approximated based on the Kraskov–Stögbauer–Grassberger (KSG) method (with parameter 4). For more information about the KSG method, the reader is referred to ~\cite{kraskov2004estimating}. Fig.\ref{fig:MI_Distortion} clearly shows that for the same level of distortion, the smart down-sampling method would leak less information about the users' power consumption compared with other methods.

\section{Concluding Remarks} \label{sec:conclusion}

In this study, we presented a privacy-preserving method for electricity consumption data, recorded by SMs, based on down-sampling or reducing the data rate of SMs data. The problem was first formulated as learning a sparse representation of the SMs time series signal which leaks minimum information about private data while keeping the reconstruction error of the original data minimum. This was implemented by simultaneously training three deep recurrent neural networks: a releaser network (providing the representation of the data to be shared), a utility network (which estimates the power consumption from the representation) and an adversary network (which attempts to infer the sensitive attribute from the representation). The performance of the proposed technique was tested based on actual SMs data and compared with a state-of-the-art algorithm, uniform down-sampling, and random down-sampling methods. The empirical results showed that this simple technique is as good as the state-of-the-art in terms of the privacy-utility trade-off, while reducing the data rate tremendously. This reduction in the temporal data resolution is of interest for smart grids since it reduces the stress on the data communication channel and relaxes the storage requirements.

\section*{Acknowledgment}

This work was supported by Hydro-Quebec, the Natural Sciences and Engineering Research Council of Canada, and McGill University in the framework of the NSERC/Hydro-Quebec Industrial Research Chair in Interactive Information Infrastructure for the Power Grid (IRCPJ406021-14). 
\bibliographystyle{ieeetr}
\bibliography{HREF}

\end{document}